\documentclass[trackchanges,twocolumn,tighten]{aastex7}
\usepackage{hyperref}
\usepackage{tabularx}
\usepackage{booktabs}
\usepackage{amsmath}
\usepackage{graphicx}
\usepackage{adjustbox}
\usepackage{subcaption}
\usepackage{float}
\usepackage{placeins}
\newcommand{\Gaia}{{\textit{Gaia} DR3}}

\newcommand\Msun{$M_{\odot}$} %

\newcommand\totmems{1172}
\newcommand\totewli{23}
\newcommand\combinedage{$418^{+32}_{-34}$\,Myr}
\newcommand\gyroage{$428\pm 93$\,Myr}
\newcommand\liage{$393.6 ^{+85.1}_{-80.9}$\,Myr}
\newcommand\evaage{$449^{+114}_{-79}$\,Myr}
\newcommand\evatot{1156}

\shortauthors{Sheffler et al.}
\begin{document}

\title{A Multi-Method Age Determination for the Ursa Major Moving Group}

\author[0009-0001-1360-8547]{Julia Sheffler}
\email[show]{jsheffler@wisc.edu}
\affiliation{Department of Physics, University of Wisconsin--Madison, 1150 University Avenue Madison, WI  53706, USA}
\affiliation{Wisconsin Center for Origins Research, University of Wisconsin--Madison, 475 N Charter St, Madison, WI 53706, USA}

\author[orcid=0009-0009-8236-0862]{Max Clark}
\affiliation{Wisconsin Center for Origins Research, University of Wisconsin--Madison, 475 N Charter St, Madison, WI 53706, USA}
\affiliation{Department of Astronomy,  University of Wisconsin--Madison, 475 N.~Charter St., Madison, WI 53706, USA}
\affiliation{Department of Astronomy, University of Michigan, Ann Arbor, MI 48109, USA}
\email{mclark.astro@gmail.com}

\author[orcid=0000-0001-7493-7419]{Melinda Soares-Furtado}
\affiliation{Department of Physics, University of Wisconsin--Madison, 1150 University Avenue Madison, WI 53706, USA}
\affiliation{Wisconsin Center for Origins Research, University of Wisconsin--Madison, 475 N Charter St, Madison, WI 53706, USA}
\affiliation{Department of Astronomy,  University of Wisconsin--Madison, 475 N.~Charter St., Madison, WI 53706, USA}
\affiliation{Department of Physics \& Kavli Institute for Astrophysics and Space Research, Massachusetts Institute of Technology, Cambridge, MA 02139, USA}
\email{mmsoares@wisc.edu}

%alphabetical authorship after the first three
\author[orcid=0009-0006-4294-6760]{Adam Distler}
\affiliation{Wisconsin Center for Origins Research, University of Wisconsin--Madison, 475 N Charter St, Madison, WI 53706, USA}
\affiliation{Department of Astronomy,  University of Wisconsin--Madison, 475 N.~Charter St., Madison, WI 53706, USA}
\affiliation{Center for Astrophysics | Harvard and Smithsonian, 60 Garden Street, Cambridge, MA 02138, USA}
\email{adam.distler@cfa.harvard.edu}  

\author[orcid=0009-0007-0488-5685]{Ritvik Sai Narayan}
\affiliation{Wisconsin Center for Origins Research, University of Wisconsin--Madison, 475 N Charter St, Madison, WI 53706, USA}
\affiliation{Department of Astronomy,  University of Wisconsin--Madison, 475 N.~Charter St., Madison, WI 53706, USA}
\email{rnarayan4@wisc.edu}  

\author[orcid=0009-0000-3122-8321]{Jenna Karcheski}
\affiliation{Department of Astronomy,  University of Wisconsin--Madison, 475 N.~Charter St., Madison, WI 53706, USA}
\affiliation{Department of Astronomy and Astrophysics, University of California, Santa Cruz, CA 95064, USA}
\email{jkarches@ucsc.edu}  

\author{Kenneth Nordsieck}
\affiliation{Department of Astronomy,  University of Wisconsin--Madison, 475 N.~Charter St., Madison, WI 53706, USA}
\email{kenneth.nordsieck@wisc.edu} 

\begin{abstract}
The Ursa Major Moving Group (UMa) is one of the closest stellar associations, yet its age has remained controversial, with published estimates ranging from 200\,Myr to 1\,Gyr. 
We present a comprehensive age analysis using the largest sample of candidate UMa members to date.
Using \textit{Gaia} DR3, we identify \totmems{} stars within 100\,pc of the Sun with 3D kinematic motions consistent with group membership.
We determine the age of UMa's dominant population using three independent methods: lithium equivalent widths (\liage{}), gyrochronology (\gyroage), and photometric variability indicators (\evaage{}).
The three methods converge on a consistent age of \combinedage{}. 
While our kinematic selection includes field stars that share UMa's space motion but are not coeval members, the convergent age determinations clearly identify a dominant population that formed together 400\,Myr ago. 
These stars are important benchmarks for studies of stellar rotation, magnetic activity evolution, and lithium depletion.
The presence of systems such as HD~63433, a young multiplanet host within the group, further illustrates the value of UMa as a laboratory for early planetary system evolution.
Our expanded catalog of kinematic candidates lays the groundwork for spectroscopic membership confirmation, refined mapping of the group's structure and chemistry, and future investigations of both stellar and planetary evolution at this key epoch.
\end{abstract}

\section{Introduction}
\label{sec:intro}
\setcounter{footnote}{0}

Young planetary systems are powerful laboratories for tracing the early evolution of planetary atmospheres, interiors, and orbital architectures. However, such systems are rare. Less than 75 confirmed exoplanet hosts have ages $<500$\,Myr with uncertainties $\leq 20$\%.\footnote{NASA Exoplanet Archive (September 2025;  \citealt{Akeson2013}).}
The Ursa Major Moving Group (hereafter UMa; also known as Collinder 285) stands out in this context as a young, nearby (nucleus distance $\approx25$\,pc) stellar moving group \citep{king03}. Among its confirmed members is the Sun-like star HD\,63433, known to host three transiting exoplanets \citep{2020AJ....160..179M, 2024AJ....167...54C}, including the nearest known Earth-sized world. Systems like HD\,63433 offer rare windows into planetary conditions analogous to Earth's early Hadean epoch, when magma oceans cooled, crusts solidified, and major volatile reservoirs were established \citep{zahnle2007}.

Refining the age and membership of UMa is therefore essential not only for stellar evolution studies but also for interpreting the properties and formation histories of young exoplanets orbiting its members.
Since the Solar System lies within the spatial extent of UMa, its members are distributed across both hemispheres and are highly accessible for photometric, spectroscopic, and interferometric follow-up. 
Despite this observational advantage, determining the age and membership of UMa members has remained challenging, with published age estimates spanning nearly an order of magnitude, from ${\sim}200$\,Myr to 1\,Gyr, as shown in Table~\ref{tab:uma_ages}. 
This ambiguity limits the utility of UMa as a benchmark for young exoplanet studies and has motivated repeated attempts to refine both the membership and age of the group.
\begin{table}[tbh!]
	\begin{center}
	\caption{Historical age estimates for the Ursa Major Moving Group, as presented in \cite{Jones15} Table~1.}
    \label{tab:uma_ages}
	\begin{tabular}{cc}
	\tableline\tableline
	Age (Myr) & Reference \\
	\tableline
	$\sim$300 	& \cite{vonhoerner_1957} \\
	300$\pm$100 	& \cite{giannuzzi_1979} \\
	630--1000 	& \cite{eggen_1992} \\
	300-400 		& \cite{soderblom93} \\
	$\sim$500 	& \cite{asiain_1999} \\
	$\sim$200 	& \cite{konig_2002} \\
	500$\pm$100 	& \cite{king03} \\
	$\sim$600 	& \cite{king_2005} \\
	393\footnote{Value corresponds to the median of the ages \citet{david_2015} report for the seven UMa stars studied in \citet{Jones15}.}	& \cite{david_2015} \\
	530 $\pm$ 40 	& \cite{brandt_2015} \\
	414 $\pm$ 23 	& \cite{Jones15} \\
	\tableline
	\end{tabular}
	\end{center} 
\end{table}

UMa has been well-studied since the 19th century, beginning with \citet{Proctor1870}, who identified a set of co-moving stars in the Ursa Major constellation heading toward a common convergent point. 
Radial-velocity confirmation soon followed \citep{Huggins1872}. \citet{Hertzsprung1909} expanded the spatial extent of the group, identifying widely separated stars sharing the same convergent motion. A major step forward towards understanding UMa's stellar population came from the doctoral work of \citet{roman49}, who assembled ${\sim}$150 members spanning A0–K3 spectral types and provided the first comprehensive description of the system.

Subsequent studies refined the membership and age using increasingly sophisticated methods. 
Combined chromospheric activity and kinematics, enabled an age estimate of ${\sim}300$\,Myr, noting lithium's potential as an age diagnostic despite substantial scatter \citet{soderblom93}. Using improved \textit{Hipparcos} \citep{Perryman1997} parallaxes and proper motions, \citet{king03} produced a conservative list of 60 high-probability members and derived an isochronal age of $500\pm100$\,Myr. 
Later spectroscopic analyses found significant chemical and chronological inhomogeneity among candidate members, with only 61\% considered chemically homogeneous despite a clear kinematic core in UVW space \citep{dopcke19}.
Bayesian isochrone fitting incorporating stellar rotation yielded an age of $530\pm40$\,Myr \citep{brandt_2015}, while interferometric measurements combined with oblate stellar models produced a more precise age of $414\pm23$\,Myr \citep{Jones15}. Notably, \citet{Jones15} also identified a ${\sim}650$\,Myr outlier (HD\,141003) within their carefully selected A-type stellar sample, further supporting contamination from kinematic interlopers.

Following the work pioneered by \textit{Hipparcos}, the European Space Agency's \textit{Gaia} mission \citep{gaiamission} provided high-precision parallaxes, proper motions, and RVs for millions of stars, enabling unprecedented 3D kinematic mapping of nearby moving groups. 
In this work, we use \textit{Gaia} Data Release 3 \citep{GaiaDR3} astrometry and radial-velocity (RV) measurements to construct a comprehensive 3D kinematic catalog of UMa members down to $G=16$\,mag. This represents the largest candidate sample assembled to date.
Our goal is not to produce a final, vetted member list, but rather to identify a broad kinematic ensemble from which a dominant coeval population can be isolated. We then constrain the age of this population using three independent stellar age indicators, which converge to yield a robust estimate.

We begin by describing our 3D kinematic identification of UMa candidates (Section~\ref{sec:kinematics}), then present lithium-based age modeling (Section~\ref{sec:lithium}), rotation period measurements and gyrochronological age estimation (Section~\ref{sec:Gyro}), and our excess variability age diagnostic (Section~\ref{sec:EVA}). We also identify a population of white dwarf candidate members (Section~\ref{sec:wdages}). Finally, we discuss our results and future prospects (Section~\ref{sec:discussion}) before summarizing our findings (Section~\ref{sec:conclusion}).

\section{Kinematic Membership Determination}
\label{sec:kinematics}

Since UMa is a dispersed moving group whose full spatial extent is uncertain, a nucleus-centered search could miss widely scattered members and underlying substructure. We therefore began with a broad \textit{Gaia} DR3 selection of stars within 100\,pc \citep{GaiaCollaboration2021}.
While core UMa members lie at $\sim$25\,pc, our 100\,pc search radius captures the full extent of high-probability members identified by \cite{king03}, the most distant of which is located at 93.35\,pc.

The \textit{Gaia} DR3 catalog provides astrometric, photometric, and RV measurements for the majority of previously established high-probability members.
To address saturation-related incompleteness for the brightest potential members, we supplemented our dataset with archival astrometry from SIMBAD (Set of Identifications, Measurements and Bibliography for Astronomical Data; \citealt{simbad}).
We added 14 UMa candidate members to the dataset, seven of which (HD\,103287, HD\,95418, HD\,129247, HD\,141003B, HD\,139006, HD\,116656, HD\,112186) were previously identified as high-probability UMa members \citep{king03}. 
Our resulting dataset included 574,545 sources.

To assess the kinematic membership of this sample, we modified the publicly-available \texttt{Python} package \texttt{FriendFinder} \citep{toffelmire2021} to search for kinematic members among all stars within our initial catalog, using HD\,115043, a well-established nucleus member of UMa \citep{king03}, as our reference star.
We then computed the expected tangential motion each star would have if it were co-moving with the UMa nucleus member. 
We identified stars whose observed tangential velocity offsets were within $\pm5$\,km\,s$^{-1}$ of these predicted values at their respective Galactocentric XYZ positions.
For reference, the high-probability nucleus members identified by \citet{king03} exhibit a maximum tangential velocity offset of 4.9\,km\,s$^{-1}$.
This resulted in a culled dataset of 15,052 proper motion candidate members.

To further filter our UMa catalog, we compared the observed RV of each proper motion candidate, as obtained from \textit{Gaia} DR3 (and SIMBAD in cases where \textit{Gaia} data was unavailable), with the RV predicted by \texttt{FriendFinder} based on kinematic consistency with UMa. 
We then removed all stars where this difference exceeded $4$\,km\,s$^{-1}$, as well as all stars fainter than $G = 16.2$\,mag (the brightness limit of available RV measurements).
For context, the maximum RV offset among the high-probability nucleus members identified by \citet{king03} is $3.5\,\mathrm{km}\,\mathrm{s}^{-1}$.
Further, RV measurements are less precise than proper motions, so applying overly tight velocity cuts risks excluding genuine members. Note that the $5\,\mathrm{km}\,\mathrm{s}^{-1}$ cut on proper motions is a stricter cut than the $4\,\mathrm{km}\,\mathrm{s}^{-1}$  cut on radial velocity since the proper motion offset accounts for two dimensions of motion.
This resulted in a culled dataset of \totmems{} 3D kinematic UMa candidates, 680 of which have both reddening and extinction corrections available. 
We note that the white dwarf proper motion candidates, which lack well-constrained RVs, were excluded from this kinematic sample---they are discussed further in Section~\ref{sec:wdages}.

\begin{figure}[h!]
\includegraphics[width=0.47\textwidth]{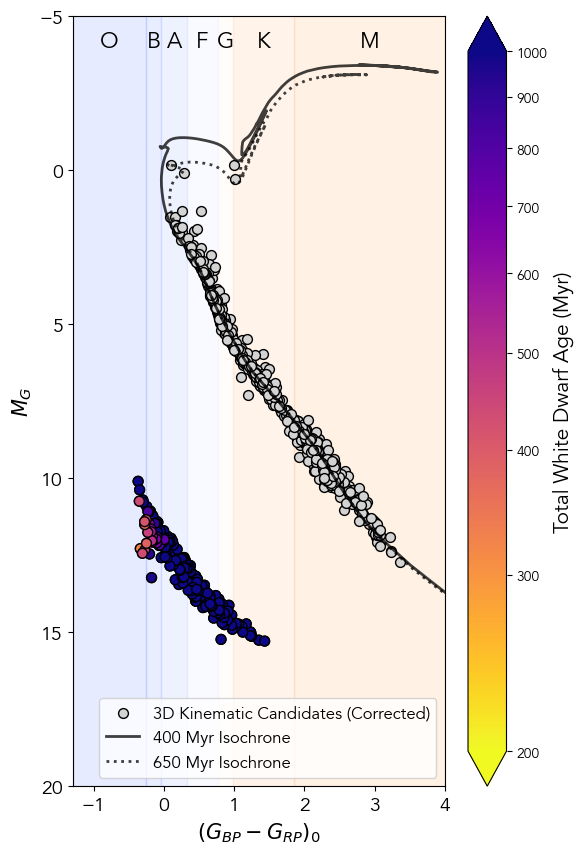}
\caption{Color–magnitude diagram for 680 reddening- and extinction-corrected 3D kinematic UMa candidates and white dwarf proper-motion candidates. The white dwarfs are colored by their \texttt{wdwarfdate} age estimate \citep{Kiman2022}. PARSEC v1.2s \citep{parsec2012} isochrones at 400\,Myr (solid black line) and 650\,Myr (dotted black line) are overplotted for comparison. Shaded background panels denote approximate spectral-type ranges, following the hexadecimal color scheme of \citet{Harre2021} for luminosity class V, subclass~5 stars.}\label{fig:cmd}
\end{figure}

We illustrate the color-magnitude diagram (CMD) positions of the 3D kinematic UMa candidates and proper motion white dwarf candidates in Figure~\ref{fig:cmd}.
Overplotted are \texttt{PARSEC} isochrones (version 1.2s, \citealt{parsec2012}) corresponding to ages of 400\,Myr (solid line) and  650\,Myr (dotted line).
While several age diagnostics suggest a dominant age near 400\,Myr \citep[e.g.,][]{soderblom93,david_2015,Jones15}, stars that are well fit by the 650\,Myr isochrone could result from systematic effects (field star contamination, photometric errors, or uncorrected stellar oblateness). They may also represent a genuine older stellar population, as suggested by the outlier star with a similar age estimate in \citet{Jones15}. The persistent age determination issues reported in prior UMa studies are immediately evident in our kinematic sample.

\begin{figure*}[t!]
\vspace*{\fill}  % pushes content to bottom of float area
\centering
\includegraphics[width=0.9\textwidth]{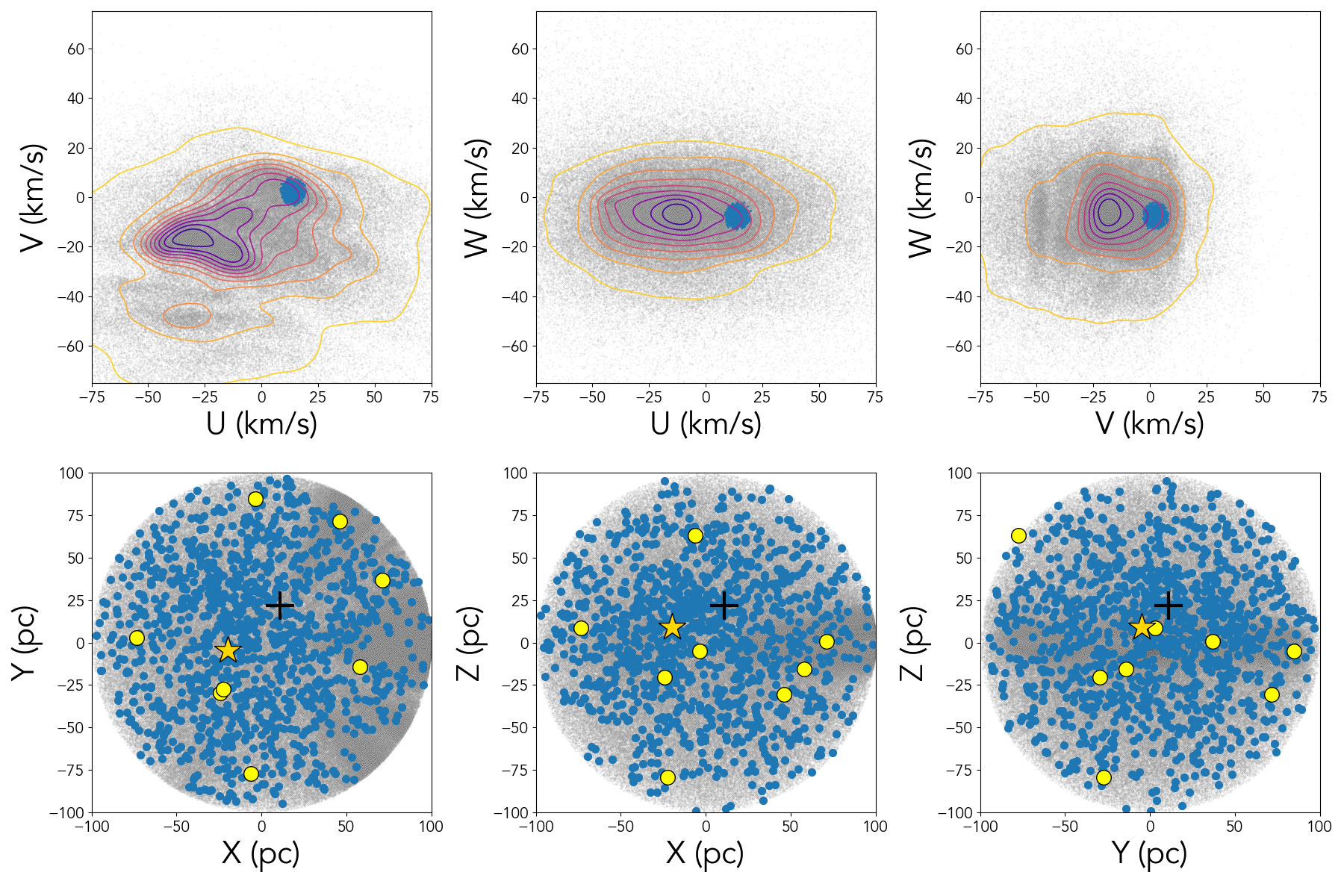}
\caption{Comparison of UMa 3D kinematic candidates (blue points) with the full sample of \textit{Gaia} stars within a 100\,pc radius (grey points). Upper panel shows Galactocentric UVW velocities with density contour lines demonstrating the over-density around UMa kinematic members.} Lower panel displays Galactocentric XYZ positions (Sun at origin). The black plus denotes HD\,115043, the core nucleus member used for comparison in our kinematic analysis.  The gold star denotes planet host HD\,63433, and yellow points denote white dwarf proper motion candidates with estimated ages of within one sigma error bounds of our age estimate as discussed later in Section \ref{sec:wdages}. 
\label{fig:placeholder2}
\end{figure*}

In Figure~\ref{fig:placeholder2}, we show our 3D kinematic UMa candidates and our full volume-limited \textit{Gaia} sample.
 In the top panel, we illustrate the Galactocentric velocity (UVW) of all stars within our initial 100\,pc search radius that have both astrometric and RV measurements (grey points).
Blue points denote stars passing our proper motion and RV membership criteria, with contours highlighting their concentration. These candidates are kinematically distinct from field stars in our 100\,pc volume.

The bottom panel shows the spatial distribution in Galactocentric XYZ coordinates centered on the Sun, with blue circles marking our 3D kinematic UMa candidates and a gold star indicating the multiplanet host HD\,63433 \citep{Mann2020,2024AJ....167...54C}. 
Yellow points represent white dwarf proper motion candidates with age and error estimates that overlap our age and error estimate for UMa which we discuss further in Section~\ref{sec:wdages}.

Notably, while kinematically distinct, the co-moving stars exhibit no clear spatial substructure and are distributed throughout the entire search volume. This uniform distribution raises concerns about potential contamination, suggesting that our kinematic selection may encompass a more complex stellar population than a single, coherent moving group.

\section{Age Determination from Lithium Equivalent Widths}
\label{sec:lithium}

To complement our kinematic analysis and address the ambiguous age of the system, we leverage lithium equivalent width (EWLi) measurements as an age diagnostic. The depletion of lithium in stellar atmospheres follows well-calibrated empirical relations with age and effective temperature \citep{Skumanich1972,soderblom1993b,eagles}, allowing us to derive both an overall ensemble age for our kinematic sample and identify individual age outliers. Stars with discrepant lithium ages are potential field contaminants or members of distinct stellar populations with different formation epochs.
We note that the observational sample will be inherently biased toward younger stars, as lithium absorption becomes increasingly difficult to detect as stars age (stars older than a few Gyr are notoriously hard to age-date; \citealt{eagles}). Despite this bias, our analysis will still help to distinguish a young, dominant subpopulation. We obtained archival EWLi measurements for our main-sequence 3D kinematic UMa candidates by querying the VizieR database \citep{VizieR2000} for published Li\,I 6707.8\AA{} equivalent width measurements. Our search returned EWLi measurements and uncertainties for \totewli{} stars, which are summarized in Table~\ref{tab:measurements} along with their corresponding references.
\begin{table*}[t!]
\centering
\begin{tabular*}{.9\textwidth}{@{\extracolsep{\fill}}cccccc}
\toprule
\textit{Gaia} DR3 ID & Teff (K) & EWli (m\AA)& e\_EWli (m\AA) & EWLi Ref. & EAGLES Age (Myr) \\
\midrule
244840902839984000 & 5860 & 54.5 & 11.4 & 1 & $803.5 ^{+3561.6}_{-739}$\\
552477109464650752 & 5480 & 79.8 & 13.5 &  1 & $416.9^{+583.1}_{-353.8}$\\
898639447612575104 & 5100 & 35.8 & 4.4 & 6 & $776.2^{+1854}_{-456.4}$\\
1092545710514654464 & 5860 & 106.4 & 3.4 & 3 & $221^{+235.8}_{-204.1}$\\
1566224665609504128 & 5930 & 94.0 & 6.0 &  6 & $281.9^{+378.9}_{-261.9}$\\
1568610571482444032 & 4830 & 46 & 5 & 6& $478.6^{+412.6}_{-298.7}$\\
1571145907856592768 & 4830 & 29.0 & 6 & 6 & $749.9^{+1268.5}_{-451.4}$\\
1571411233756646656 & 4830 & 21.8 & 2.6 & 6 &$988.6^{+1862.5}_{-595.0}$\\
1870103394346059264 & 6050 & 98 & 15 & 7 & $190.5^{+433}_{-176.3}$\\
3285218186904332288 & 6050 & 88.0 & 2.1 & 6 &$272.3^{+504.0}_{-253.2}$\\
3399063235755057792 & 5860 & 103.6 & 4.4 & 6 & $239.9^{+255.6}_{-221.7}$\\
5450107207053038592 & 4300 & 50 & 10& 4 & $251.2^{+211.2}_{-156.8}$\\
5808612830236138368 & 6050 & 116.1 & 3.1 &  3 &$118.9^{+186.6}_{-108.0}$\\
6519711873737172608 & 5990 & 85.0 & 10.0 &  5 &$335.0^{+724.3}_{-313.1}$\\
\hline
\hline
358549337366815872 & 6050 & 47.1 & 9.3 &  1 & $>$4.3\\
723944775287648256 & 5770 & 11.6 & 2.3$^*$ & 2 & $>$5.7\\
1651645517813056384 & 5990 & 49.3 & 9.9$^*$ &  1 &$>$4.1\\
3345968334645285376 & 5270 & 13.2 & 0.8 & 8&$>467.7$\\
3381727613874753536 & 5990 & 15.3 & 3.4 & 5  & $>4.2$\\
3534658789261176064 & 5930 & 110.0 & 22.0$^*$ &  4 & $179.9^{+351}_{-166.6}$\\
4038724053986441856 & 5930 & 100.0 & 20.0$^*$ &  4 & $239.9^{+501.4}_{-223.5}$\\
4281201414717499520 & 6050& 10 & 2$^*$& 5 & $>3.4$ \\
5173902189571919872 & 6280& 112.8 & 22.6$^*$& 9 &  $49.0^{+242.8}_{-43.3}$\\
\bottomrule
\end{tabular*}
\caption{Lithium equivalent width and effective temperature measurements for 23 main-sequence UMa 3D kinematic UMa candidates. Effective temperatures are determined using reddening-corrected \Gaia{} GBP-GRP values and the relations presented in \citet{Mamajek}. Lithium equivalent width reference key: 1) \citet{Guillout2009}, 2) \citet{Lubin2024}, 3) \citet{Maldonado2010}, 4) \citet{torres2006}, 5) \citet{Cutispoto2002}, 6) \citet{AmmlervonEiff2009}, 7) \citet{Frasca2018}, 8) \citet{lopezsantiago2010}, 9) \citet{chen2001}. The superscript $*$ symbol indicates that an error estimate was not provided in the literature, in which case we imposed a conservative 20\% error. Those under the double line were excluded from the cluster analysis due to unreliable errors or unbound eagles age estimates.}
\label{tab:measurements}
\end{table*}
We estimated effective temperatures for each star by linearly interpolating reddening-corrected \textit{Gaia} color ($G_{\mathrm{BP}}-G_{\mathrm{RP}}$) values using the Modern Mean Dwarf Stellar Color and Effective Temperature Sequence digital table originally provided in \cite{Mamajek}.\footnote{\url{https://www.pas.rochester.edu/~emamajek/EEM_dwarf_UBVIJHK_colors_Teff.txt}} 
The resulting effective temperatures are listed in Table~\ref{tab:measurements}.
We adopted a conservative effective temperature uncertainty of 200\,K for all targets to account for uncertainties in the color-temperature calibration, reddening corrections, and photometric measurements.

To perform this analysis, we use the EAGLES (Estimating Ages from Lithium Equivalent Widths) software version 2.0 \citep{eagles,Weaver2024}.
The model characterizes the relationship between EWLi (and its intrinsic dispersion), effective temperature, and age, enabling both individual stellar ages and ensemble age estimates for coeval populations.
This tool employs empirical models calibrated on approximately 6,000 stars across 52 open clusters from the \textit{Gaia}-ESO survey, providing age estimates for stars with effective temperatures between 3000--6500\,K  and metallicities between $-0.3 < \textrm{[Fe/H]} < 0.2$\,dex. 
The updated version incorporates an artificial neural network model that improves lithium depletion modeling, treatment of the lithium dip feature \citep{Boesgaard1986}, and dispersion estimates. Age-dating capabilities are constrained between 2\,Myr--1\,Gyr.
Stars with discrepant individual ages that don't match the main population will be apparent as outliers in the analysis.

\begin{figure}[]
\includegraphics[width=0.47\textwidth]{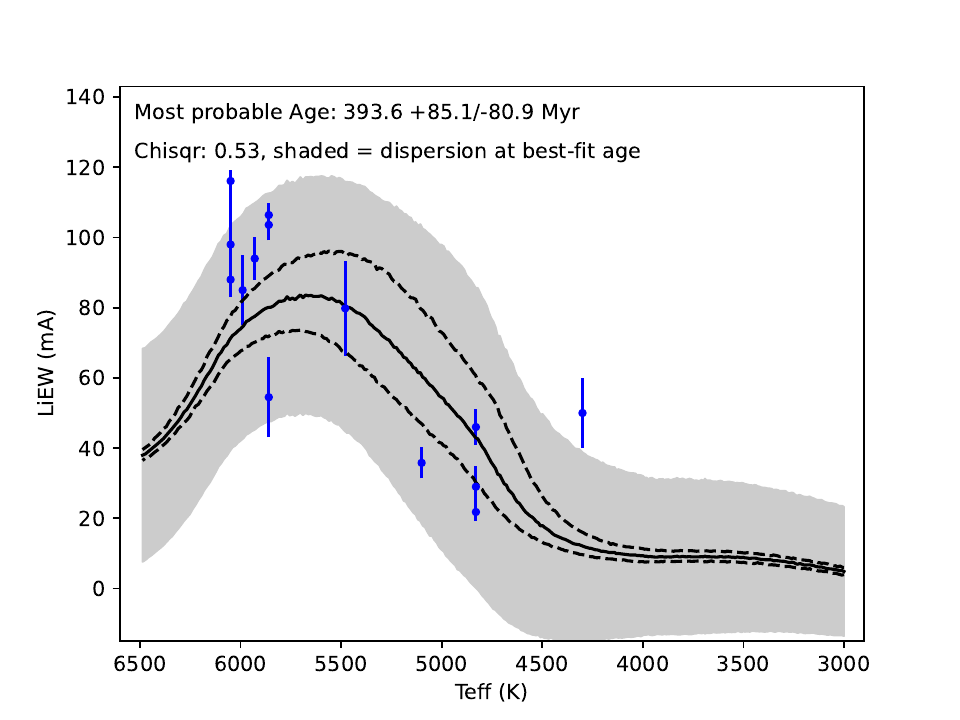}
\caption{EAGLES model output illustrating the lithium equivalent width measurements versus effective temperature for 14 UMa kinematic members. The grey shaded region shows the intrinsic dispersion in lithium equivalent width at the most probable ensemble age of \liage{}.}
\label{fig:ewli}
\end{figure}

The results of our lithium age analysis are shown in Figure~\ref{fig:ewli}. 
We  list the individual most probable lithium-based ages and associated errors for each of the 23 stars in Table~\ref{tab:measurements}. For six of these 23 stars, the archival source did not report an error estimate. Another 3 stars produced unbounded age estimates in eagles as shown in the table. All 9 of these stars are excluded from the group membership calculation and listed bellow the double line in Table~\ref{tab:measurements}.

We run EAGLES cluster fitting code to our remaining 14 reliable EWLI sources. The EAGLES code version 2.0 uses a neural network to fit isochrones to our data. Figure~\ref{fig:ewli} shows the best fit isochrones and the most-probable moving group age of \liage{}, with a low dispersion (grey-shaded region).
Further, $\chi^2=0.53$ indicates a robust age estimate for this sample, suggesting the lithium measurements are consistent with a single coeval population. 
These results are consistent with UMa's status as a relatively young moving group, independently validating the UMa age estimate determined by \citet{Jones15}.

\section{Age Determination from Gyrochronology}\label{sec:Gyro} 
To provide an independent age diagnostic that complements our lithium equivalent width analysis, we employ gyrochronology. This method enables the determination of stellar ages from rotation periods and stellar masses based on the empirical relationship first established by \citet{Skumanich1972}.
While traditional gyrochronology relied on a simple relationship between stellar rotation, mass, and age, observations reveal deviations that complicate age determination. To address these complexities, we use the open-source tool \texttt{gyro-interp} \citep{2023ApJ...947L...3B}, which employs a data-driven approach calibrated on observational data from 11 open clusters with well-defined age constraints. 
By interpolating between these cluster datasets, \texttt{gyro-interp} produces posterior age distributions for individual main-sequence stars based on their rotation periods and effective temperatures.
The tool is amenable to stars with effective temperatures between 3800-6200\,K and ages between 0.08-4\,Gyr.

\begin{figure*}[t]
\centering
\includegraphics[width=0.97\textwidth]{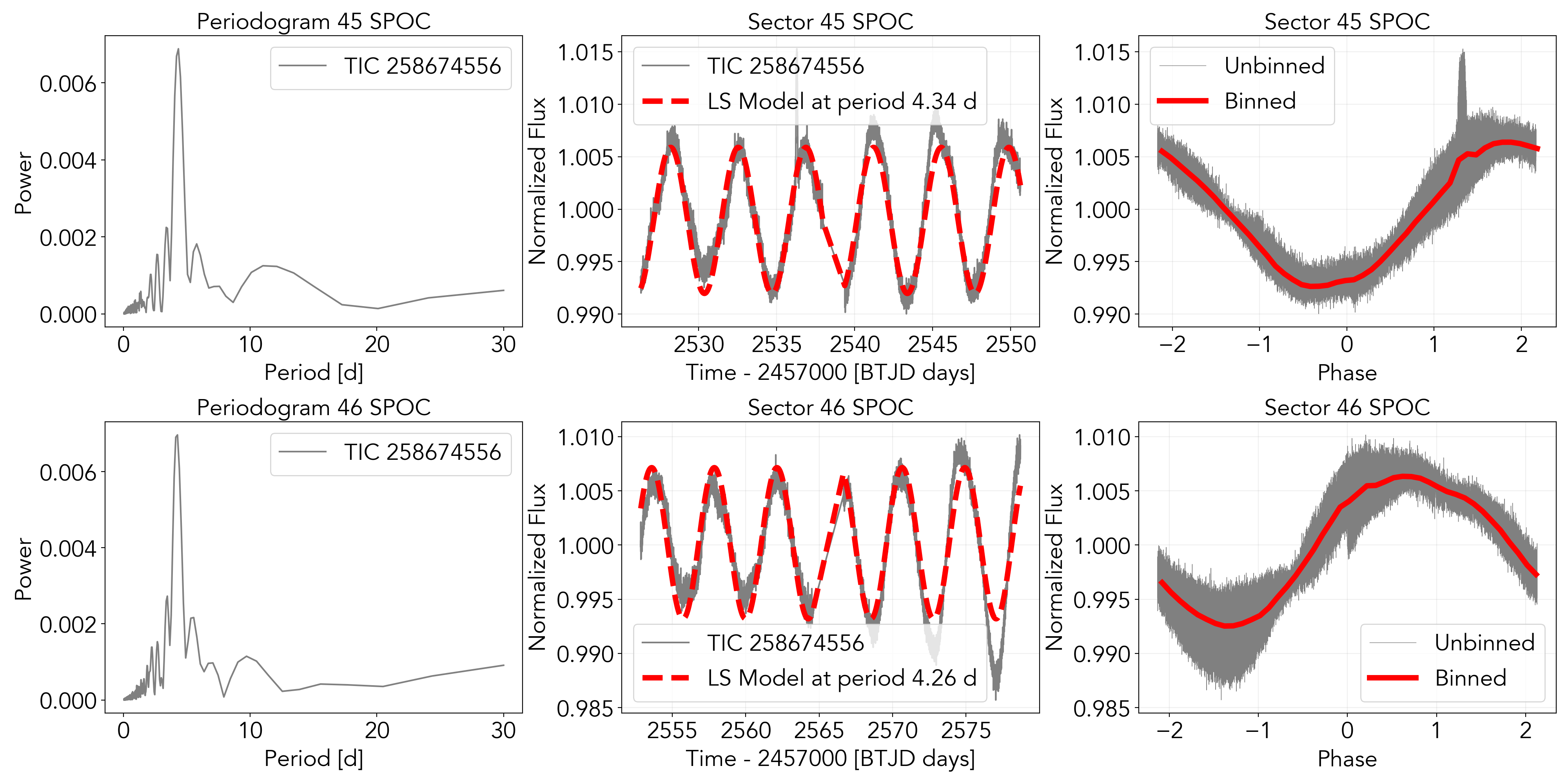}
\caption{Example TESS rotation period analysis for the target TIC~258674556. Top row: SPOC Sector 45 results, showing the Lomb–Scargle periodogram (left), the corresponding phase-folded light curve with the best-fit sinusoid (middle), and the normalized light curve with binned points overlaid (right). The dominant periodogram peak occurs at 4.34 days. Bottom row: SPOC Sector 46 results shown in the same format, with a dominant peak at 4.26 days.}\label{fig:LC}
\end{figure*}

\subsection{Measuring Photometric Rotation Periods}
\label{subsec:rot_data}

The availability of high-precision time series photometry from the Transiting Exoplanet Survey Satellite (TESS; \citealt{Ricker2015}) enables gyrochronological age determination for substantially more UMa members than our lithium sample. Rotation periods are detected by the periodic brightness modulations caused by starspots rotating in and out of view. However, this method once again introduces a selection effect toward younger stars, as older field stars often rotate too slowly for reliable period detection within TESS' nominal 27-day observing baseline. Nevertheless, this approach can still test whether a dominant young population exists in our 3D kinematic population.

Not all UMa 3D kinematic UMa candidates are suitable for gyrochronological analysis. 
Despite the fact that the sample consists of stars in the Solar neighborhood, local dust clouds, molecular clouds, and filamentary structures can cause significant reddening even at short distances.
More specifically, we observe \textit{Gaia} DR3 reddening values for members of our 3D kinematic sample as large as 0.85\,mag.
For context, at BP–RP = 0.85\,mag, a 0.85\,mag correction corresponds to a temperature shift of approximately 4,000\,K, while a 0.2\,mag offset corresponds to a temperature shift of approximately 700\,K.
Therefore, to ensure accurate placement in rotation-temperature space, we omit targets lacking this correction measurement. 

To measure rotation periods, we employed the \texttt{Lightkurve} software package, a standard tool for TESS photometry analysis \citep{lightkurve}.
This included identifying and downloading existing data from the Barbara A.~Mikulski Archive for Space Telescopes (MAST) archive \citep{MAST}.
Of our initial \totmems{} 3D kinematic UMa candidates with \textit{Gaia} DR3 correction values, 682 stars had available TESS light curves from one of the following three data reduction pipelines: Science Processing Operations Center (SPOC; \citealt{jenkinsSPOC2016}), TESS Light Curves From Full Frame Images (TESS-SPOC; \citealt{2020RNAAS...4..201C,tess-spoc}), or the Quick Look Pipeline (QLP) pipelines \citep{QLP, QLP2,qlp2020}. 
We then applied an iterative sigma clipping tool to remove outliers in the light curve data more than $3{\sigma}$ away from the median.  
Using the Lomb-Scargle periodogram \citep{lomb_scargle_periodogram}, we identified stellar rotation periods. This algorithm is well-suited for detecting near-sinusoidal signatures induced by star spot modulations.
We searched for periodic signals in the range of 0.04--30 days.
After performing the periodogram analysis, we generated diagnostic plots for each available TESS sector and pipeline, including the power spectrum, the time series with a sinusoidal model overlaid at the peak period, and the phase-folded light curve constructed using the strongest periodogram peak.
We illustrate an example diagnostic plot in Figure~\ref{fig:LC}.

Two team members independently vetted each source by eye. We excluded those for which no consistent period was found across the available sectors. 
If harmonics caused conflicting periods, we prioritized the longer period, as spot-driven variability is often symmetric, causing periodograms to return half the true rotation period. 
When no clear harmonic was present, we chose the period with the most sinusoidal phase-folded light curve. Sources without a confident consensus from both reviewers were removed. 

We used all available TESS reduction pipelines. In most cases, the period estimates agreed for a given sector. For period reporting, we adopted the SPOC pipeline whenever the data were available and the resulting phase-folded light curves revealed a clear repeating signal. SPOC apertures typically minimize contamination while maximizing signal-to-noise by incorporating sophisticated systematic corrections, including cotrending basis vectors. This preserves intrinsic stellar variability after removing systematics. Otherwise, we selected the pipeline whose periodogram and phase-folded light curve provided the most convincing and coherent rotational signal.

\begin{table*}[]
\centering
\caption{The 15 brightest 3D-kinematic candidate members with reliably vetted rotation periods}
\label{tab:rot_period}
\begin{tabular}{llcccccccc}
\toprule
\textit{Gaia} DR3 ID & TIC ID & RA & DEC & BP-RP  & \textit{Gaia} mag & $P_{\mathrm{rot}}$ & e$P_{\mathrm{rot}}$ & Source \\
& & (degrees) & (degrees) & (mag)  & (mag) & (days) & (days) & \\
\midrule
1092545710514654464 & 417762326 & 129.80 & 65.02 & 0.80 & 5.50 & 4.98 & 1.00 & QLP\\
4038724053986441856 & 329574145 & 271.60 & -36.02 & 0.78 & 5.81 & 6.02 & 1.20  & SPOC\\
6472858766996632704 & 422044019 & 301.90 & -55.02 & 0.70 & 6.13 & 3.17 & 0.63  & SPOC\\
3886447534566460032 & 160271830 & 159.73 & 16.13 & 0.55 & 6.51 & 3.23 & 0.65  & TESS-SPOC\\
3345968334645285376 & 415563103 & 91.67 & 15.54 & 0.98 & 6.54 & 8.04 & 1.61  & SPOC\\
4281201414717499520 & 227409000 & 284.15 & 4.26 & 0.73 & 6.58 & 3.96 & 0.79  & SPOC\\
1785414129672353408 & 60446208 & 319.80 & 17.82 & 0.51  & 6.62 & 1.25 & 0.25  & SPOC\\
875071278432954240 & 130181866 & 117.48 & 27.36 & 0.85 & 6.74 & 6.45 & 1.29  & SPOC\\
2583819069342416384 & 408261976 & 18.08 & 12.28 & 0.65 & 6.77 & 4.09 & 0.82  & SPOC\\
3285218255623808640 & 283792891 & 63.86 & 6.20 & 0.86  & 6.78 & 6.50 & 1.30  & SPOC\\
2328281189678351232 & 33982876 & 357.33 & -27.85 & 0.66  & 6.83 & 3.87 & 0.77  & SPOC\\
6519711873737172608 & 121421711 & 341.86 & -44.97 & 0.76  & 7.09 & 5.31 & 1.06  & SPOC\\
2091285551421214976 & 68266718 & 280.09 & 33.68 & 0.62 & 7.17 & 2.59 & 0.52  & SPOC\\
3086002004397666304 & 271310754 & 117.24 & 0.66 & 0.43 & 7.39 & 0.68 & 0.14  & SPOC\\
278914871261809920 & 252221650 & 72.05 & 59.24 & 0.41  & 7.42 & 0.08 & 0.02  & SPOC\\
\bottomrule
\end{tabular}
\end{table*}

Our vetting procedure resulted in 147 rotation periods, the 15 brightest of which are listed in Table~\ref{tab:rot_period}. For each star, the reported rotation period is the median across all sectors whose phase-folded light curves show adequate phase coverage and low photometric scatter. This approach filters out problematic single-sector results and yields a robust period estimate for each target. 
The period uncertainty reported in Table~\ref{tab:rot_period} is $20\%$ of the measured period.
\citet{Reinhold_2015} found that multiple significant periodogram peaks attributed to differential rotation are common among active stars, allowing period deviations of 10--20\%.
We therefore conservatively apply a 20\% uncertainty on our reported rotation periods. 

\subsection{Rotational Age Estimation}
Before conducting our age analysis with \texttt{gyro-interp}, we applied additional temperature cuts to remain within the valid parameter space of the gyrochronology models. As in Section~\ref{sec:lithium}, we determined effective temperature by interpolating across The Modern Mean Dwarf Stellar Color and Effective Temperature Sequence digital table \citep{Mamajek}.

We adopted the upper and lower temperature bounds for \texttt{gyro-interp} of 6200K and 3800K. The temperature ceiling of $6200$\,K, ensures that we removed all sources above the Kraft break (at which point we do not expect to see surface spot modulation). We impose a temperature floor of $3800$\,K, as lower-mass stars show wide period dispersion that makes the gyrochronology sequence ambiguous below this threshold.

\begin{figure*}[tbh!]
    \centering
    \begin{subfigure}{0.8\textwidth}
        \includegraphics[width=\textwidth]{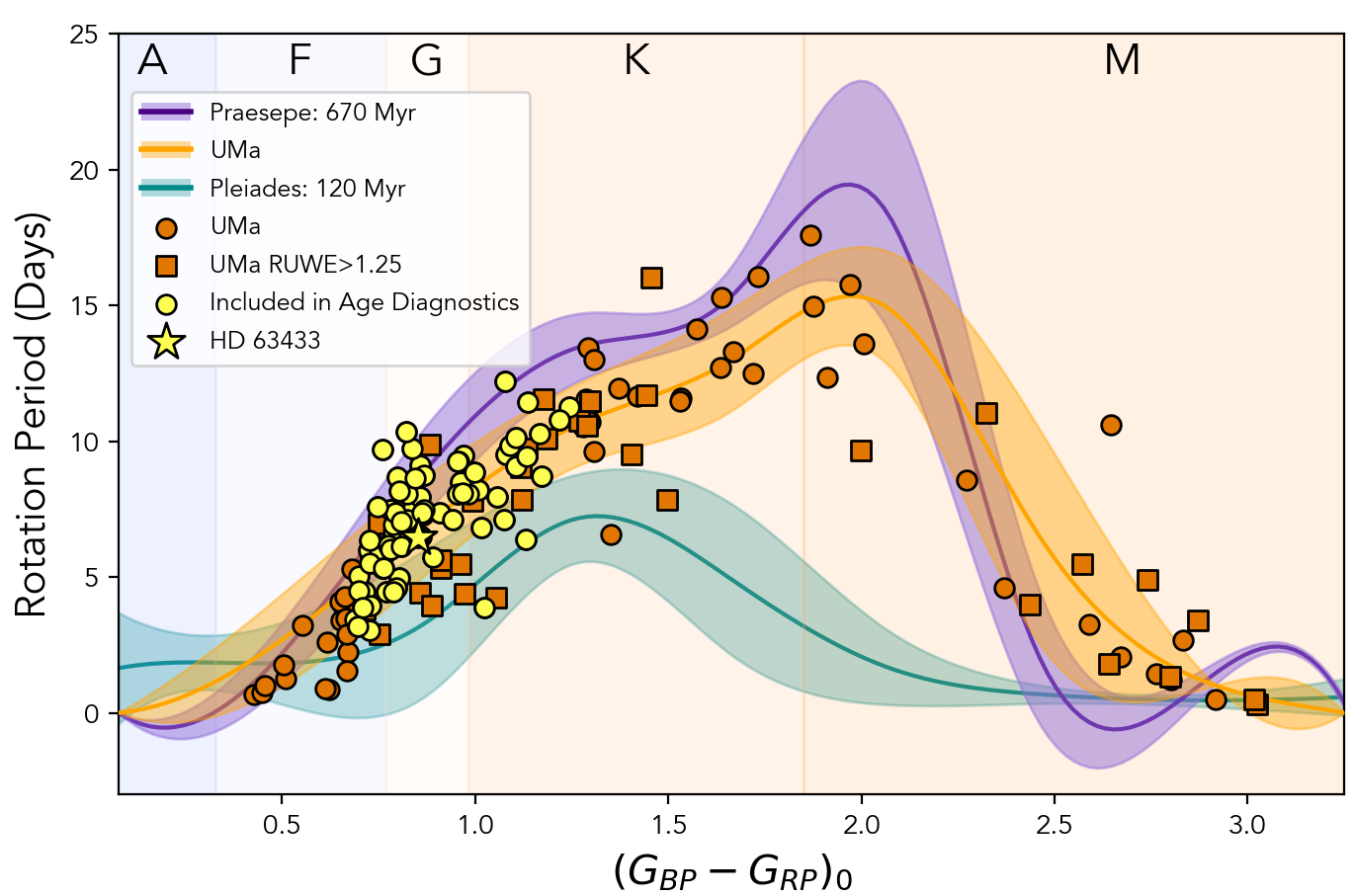}
    \end{subfigure}
    \vspace{-4.5mm}
    \begin{subfigure}{0.97\textwidth}
        \includegraphics[width=\textwidth]{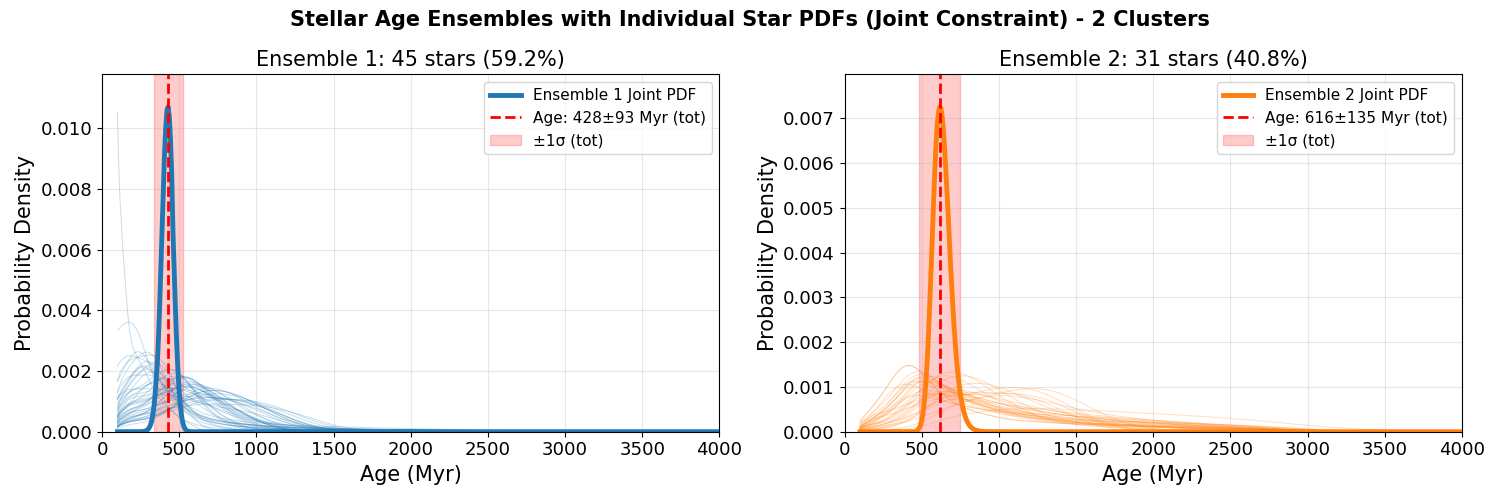}
    \end{subfigure}
    \vspace{-4.5mm}
    \begin{subfigure}{0.97\textwidth}
        \includegraphics[width=\textwidth]{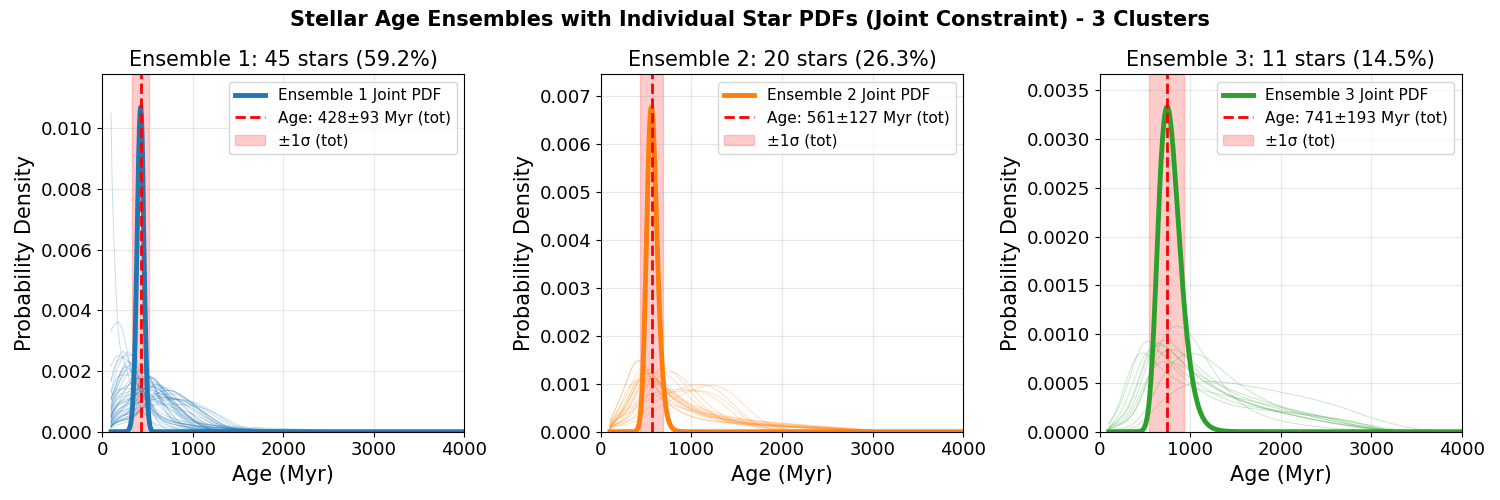}
    \end{subfigure}    
    \caption{\textit{Top:} Color–rotation diagram for UMa 3D kinematic candidate members with TESS-derived rotation periods. Yellow points denote sources included in our age analysis; the yellow star indicates HD\,63433. Orange squares and circles indicate excluded stars (RUWE cuts and temperature/color cuts, respectively). Shaded sequences show the median $\pm$1 
    median absolute deviation envelopes for the Pleiades (120 Myr; \citealt{Rebull16}), Praesepe (670 Myr; \citealt{Douglas19}), and UMa. \textit{Middle and bottom:} Adaptive clustering results for two- and three-cluster cases, respectively. The dominant 428\,Myr population remains consistent despite changes in the older clustering.}
    \label{fig:col-rot-clustering}
\end{figure*}

Further, since blended binaries are known to skew age-rotation correlations, we incorporated a \textit{Gaia} DR3 Re-normalised Unit Weight Error (RUWE) threshold of $<1.25$, a well-established metric to indicate potential binarity \citep{Penoyre2022}. This reduced our sample to 97 sources that met our selection criteria. Of these 97 stars, we found that many redder stars had much broader posterior distributions in age, and were less reliable for age estimates. As a result, we further restrict the sample to FGK stars less than 1.25 in the BP-RP mag \textit{Gaia} band pass (corrected for reddening and extinction). In the top panel of Figure~\ref{fig:col-rot-clustering}, the $76$ sources that passed all selection cuts and were included in our age estimation are plotted in yellow. We depict the single stars with measured rotation periods that were outside the temperature or color cuts, as well as stars with $\textrm{RUWE}>$1.25 indicated by squares, in orange. 

Most UMa stars fall between the gyrochronology sequences of the Pleiades (120\,Myr; \citealt{Rebull16}) and Praesepe (670\,Myr; \citealt{Douglas19}), consistent with an intermediate age. This placement aligns well with the previously inferred age of the UMa group from \cite{Jones15}.
Notably, we observe significant scatter in the rotation periods, particularly among the G- and K-type stars, indicating potential contamination among our 3D kinematic candidate members. 
Since such contaminants primarily increase dispersion without shifting the underlying coeval sequence, the dominant population in the color–rotation plane remains clearly identifiable.

Once equipped with the vetted period estimates and effective temperatures for our sample of stars, we used \texttt{gyro-interp} to estimate posterior age distributions for each star individually. More specifically, we computed the posterior age distributions across a linear age grid ranging between 100\,Myr to 4\,Gyr with 500 points. We observed a wide spread in the corresponding shapes of the posterior age distributions, with fast-rotating stars peaking at 100\,Myr, and others that peak above 1\,Gyr, suggesting that our candidate catalog likely contains a significant contaminating population. To address the contamination, we apply a spectral clustering algorithm to isolate the age of the dominant population and estimate contamination rates.

Spectral clustering \citep{vonLuxburg2007} is particularly well-suited for identifying groups in complex, non-Gaussian distributions, including complex distributions with irregular boundaries that traditional methods would miss. 
The algorithm groups stars based on the similarity of their age probability distributions, effectively separating the narrow, well-defined age peak of coeval members from the broad, scattered distributions of field contaminants. We implemented this using \texttt{scikit-learn}'s \texttt{SpectralClustering} \citep{Pedregosa2011} with $k=2$ and $k=3$ clusters. 

In the middle and bottom panels of Figure~\ref{fig:col-rot-clustering} we illustrate the ensemble of these posterior age distributions for our UMa candidate members, fitting for two and three clusters (top and bottom panels, respectively). In both cases, we find a dominant population with a rotational age estimate of \gyroage. Of the initial 76 stars included in the age diagnostic, 45 (59\%) belong to the dominant coeval population, while 31 fall outside, indicating a contamination rate of approximately $41\%$. Our contamination percentage is in good agreement with \cite{dopcke19}, where $14/23$ sources were determined to be confirmed members, resulting in a contamination rate of $39\%$.

%\begin{figure*}[tbh!]
%\includegraphics[width=.95\textwidth]{UMa_Rotation_Adaptive_clustering_2_clusters.png}
\section{Excess Variability Based-Age (EVA)}
\label{sec:EVA}
Stellar variability is a well-known indicator of youth. While individual stellar variability measurements can be widely scattered, considering moving groups, clusters, or associations allows for greater precision. The method for approximating the stellar variability-age relation \texttt{EVA} presented in \citet{Barber23} uses regularly measured parameters in the \textit{Gaia} catalog, to which all of our stars belong. 
More specifically, the tool leverages the excess photometric scatter in the $G$, $BP$, and $RP$ \textit{Gaia} bands. By calibrating this excess variability against clusters of known ages, the method derives an empirical age-variability relation that can be applied to field stars and moving group candidates. The tool achieves ages within 10\%–20\% of the true value for associations younger than 2.5\,Gyr.

\begin{figure}[h!]
\includegraphics[width=0.47\textwidth]{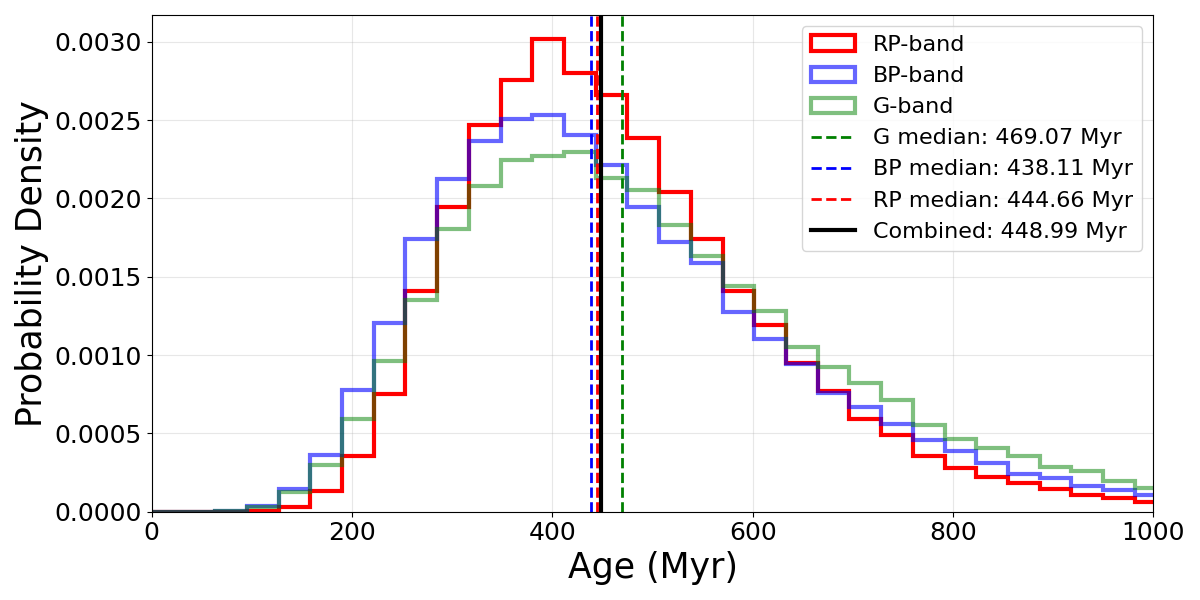}
\caption{Age probability distribution, as derived via the \texttt{EVA} age analysis, which uses three \textit{Gaia} bands. The distributions show results from $G$-band (green), $BP$-band (blue), and $RP$-band (red) rotation periods, with vertical dashed lines indicating median ages. The combined age distribution (black solid line) yields a median age of 449.3\,Myr. The well-defined peaks indicate good agreement between independent photometric bands, supporting a dominant coeval stellar population.}
\label{fig:eva}
\end{figure}

Of the \totmems{} 3D kinematic candidate members \evatot{} have the appropriate data available allowing for a much larger data set than in previous two age estimates. \texttt{EVA} was tested for reliability for stars within a \textit{Gaia} color magnitude range of $1<G_{BP}-G_{RP}<2.5$, so we reduced our sample down to 539 after cuts on color. 
While less precise than the previous two methods, \texttt{EVA} allows us to include more than half the candidate members in the analysis, reducing any bias in the smaller data samples. The \texttt{EVA} code produces a histogram with an age estimate of \evaage{} as shown in Figure~\ref{fig:eva}.

\section{White Dwarf Proper Motion Candidates}\label{sec:wdages}
We identified 225 candidate white dwarf UMa members among our proper motion–selected sample.
Unlike main sequence stars, RV measurements cannot effectively constrain white dwarf membership due to large systematic uncertainties introduced by gravitational redshift effects. The strong surface gravity of white dwarfs ($\log{g}\sim8$\,dex) produces gravitational redshifts that exceed the intrinsic velocity dispersion of the moving group. 

We therefore employed white dwarf cooling ages as the primary additional membership criterion to distinguish likely UMa members from field interlopers.
For this analysis, we use \texttt{wdwarfdate} \citep{Kiman2022}, an open-source tool that derives white dwarf age estimates from effective temperatures and surface gravity measurements. The tool utilizes a Bayesian framework with MCMC sampling to compute the total age of a source by combining empirical white dwarf cooling models \citep{bedard20} and the estimated lifetimes of their main-sequence progenitors based on models of the initial-final mass relation  \citep{choi16,dotter16,cummings18}.

The white dwarf proper motion candidates were determined by cross-matching our total proper motion sample with the \citet{Vincent24} catalog of white dwarfs, allowing us to obtain parameters required for age estimation. Using the effective temperatures, surface gravities, and uncertainties from this catalog, we computed ages with \texttt{wdwarfdate}, employing the \texttt{PARSEC} initial-final mass relation \citep{cummings18} and the DA or non-DA atmospheric models based on the spectral classifications provided in the catalog. 

\begin{figure}[t!]
\centering
\includegraphics[width=0.48\textwidth]{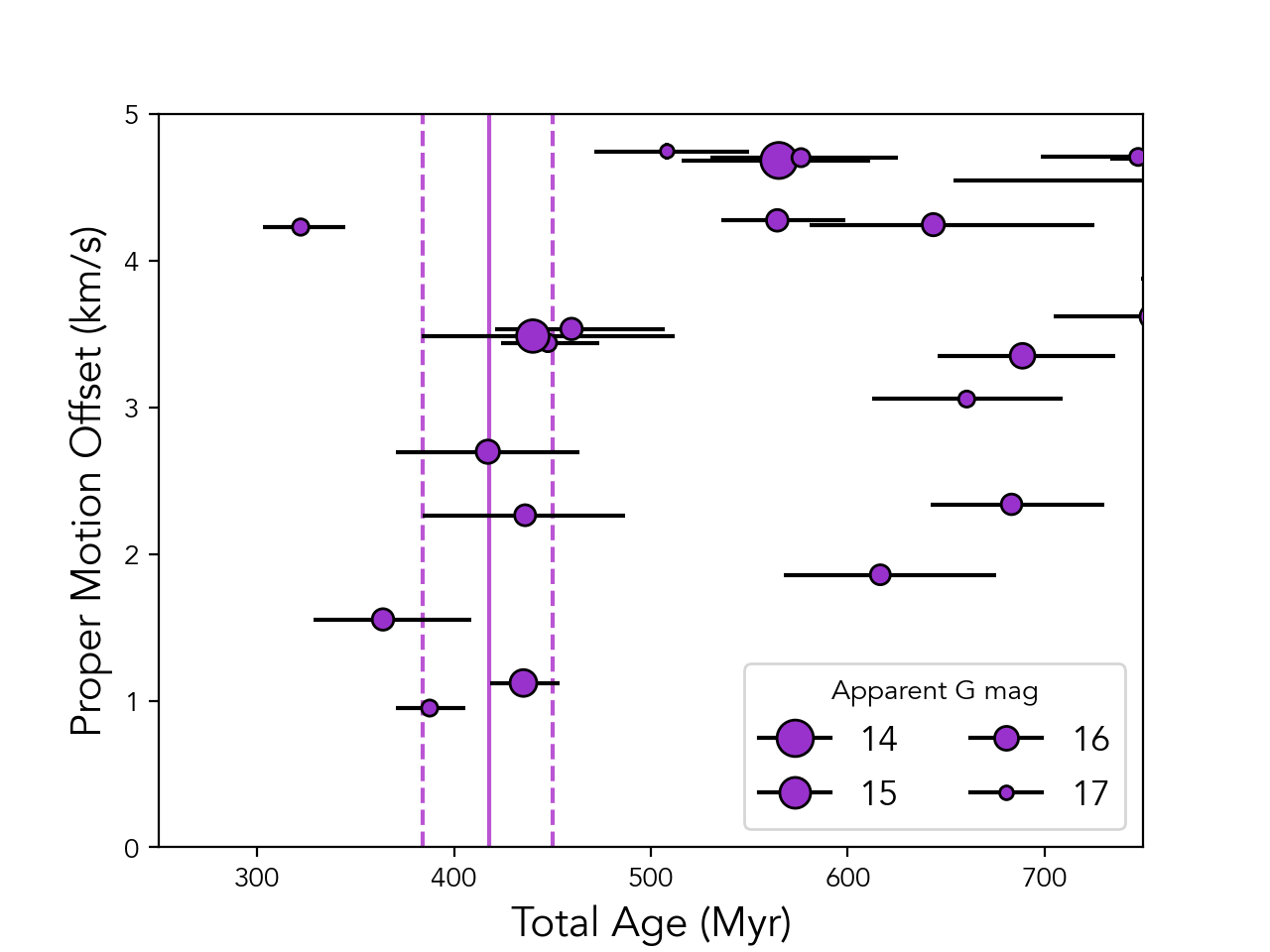}
\caption{Total age versus proper motion offset for UMa white dwarf candidates. Ages are derived from effective temperatures and surface gravities in \citet{Vincent24}; proper motion errors are smaller than point sizes. Symbol sizes scale with apparent \textit{Gaia} magnitude, highlighting strong candidates for spectroscopic follow-up. Vertical dashed lines mark the constraints of our joint age  estimate (\combinedage{}).}\label{fig:wdages}
\end{figure}

We identified likely UMa WD members by filtering the white dwarf proper motion candidates for ages that overlapped with our dominant component age estimate of \combinedage{}.  
Figure~\ref{fig:wdages} illustrates the ages, proper motion offsets, and apparent magnitude (indicated by size) of each of the eight white dwarfs within the bounds of our joint age estimation of \combinedage{} and errors. 

These targets and their corresponding proper motion offsets, apparent magnitudes, effective temperatures, surface gravities, total ages, and spectral types are listed in Table~\ref{tab:wd_ages}.
\begin{table*}[]
    \centering
    \begin{tabular}{lccccccc}
        \toprule
        \textit{Gaia} DR3 ID & PM Offset & Apparent g mag & Teff & logg & Cooling Age & Total Age & Spectral Type \\
         & (\,km\,s$^{-1}$) & (mag) & (K) & (dex) & (Gyr) & (Gyr) &\\
        \midrule
        1751777801234838144 & 1.55 & 16.29 & 20553 & 8.48 & 0.18 & 0.36 & DA \\
        1977625873372742016 & 2.70 & 16.06 & 20070 & 8.40 & 0.17 & 0.42 & DA \\
        2908425653829891712 & 1.12 & 15.61 & 23424 & 9.07 & 0.39 & 0.44 & DA \\
        3452373568124842752 & 3.44 & 16.68 & 21047 & 8.93 & 0.40 & 0.45 & DA \\
        3557764648160770176 & 2.26 & 16.34 & 20278 & 8.37 & 0.15 & 0.44 & DA \\
        4256698798129293568 & 3.54 & 16.29 & 17931 & 8.48 & 0.27 & 0.46 & DA \\
        5063894054052370304 & 0.95 & 16.81 & 20884 & 8.81 & 0.32 & 0.39 & DA \\
        6706768660234603136 & 3.49 & 14.71 & 25231 & 8.26 & 0.05 & 0.44 & DA \\
        \bottomrule
        \end{tabular}
    \caption{White dwarfs with proper motion agreement to well-established UMa nucleus member HD\,115043. The total age of each of these targets is consistent (within errorbars) to our age estimate of \combinedage{}.}
    \label{tab:wd_ages}
\end{table*}
This population of eight white dwarf candidate members represent valuable probes of intermediate-mass stellar evolution. With progenitor masses of 2--4\,\Msun{} (based on the \combinedage{} age estimate), these white dwarfs provide empirical constraints on main-sequence lifetimes, mass-loss rates, and initial-final mass relations. Their proximity and known age make them ideal targets for detailed spectroscopic follow-up to study white dwarf atmospheric composition and cooling physics.

\section{Discussion}
\label{sec:discussion}
\begin{figure*}[]
\centering
\includegraphics[width=0.98\textwidth]{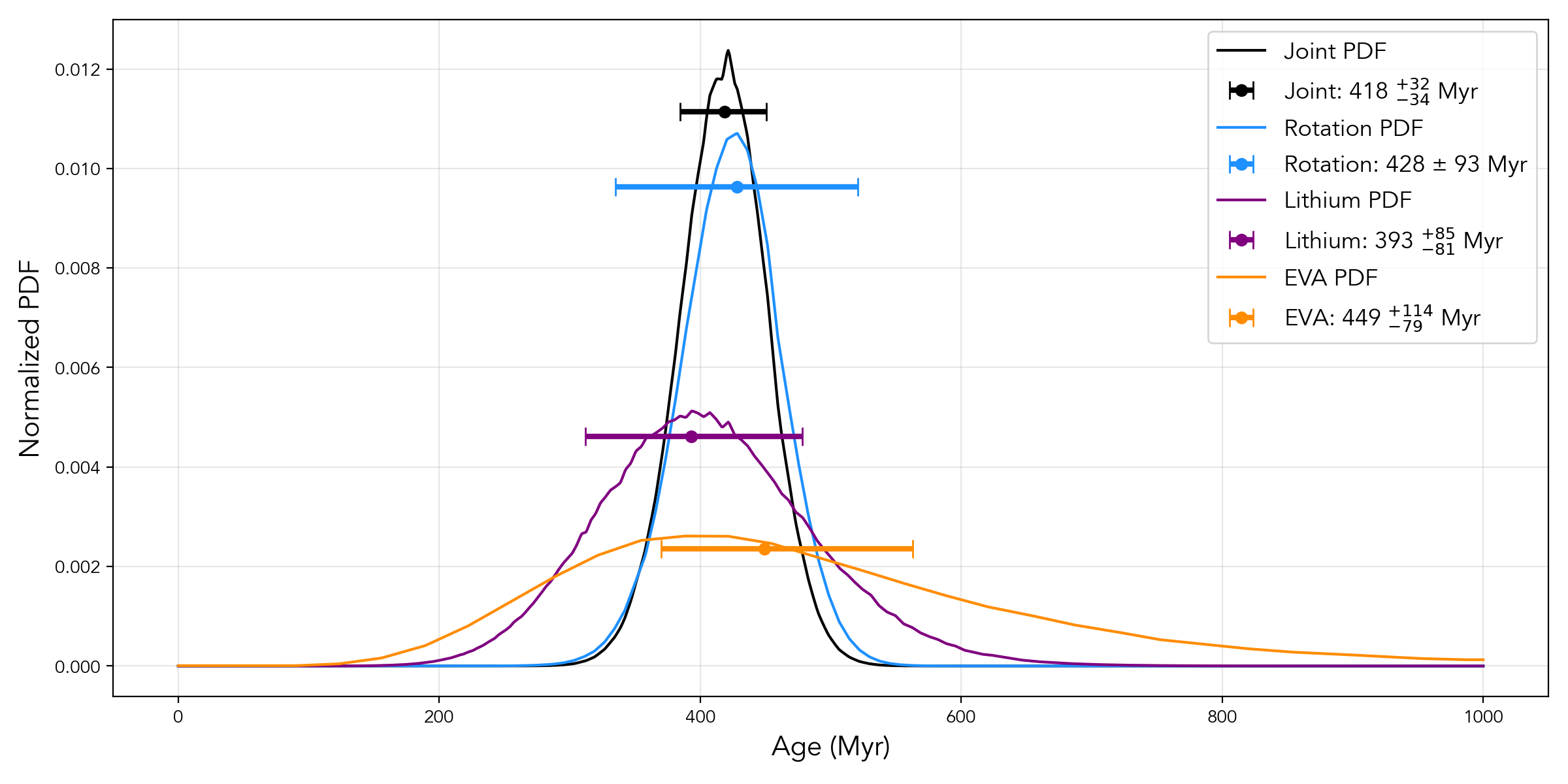}
\caption{Gyrochronological age (blue), lithium-temperature relation age (purple), photometric excess age (\texttt{EVA}, orange), and joint posterior (black) with mean and error bounds. All three methods converge on an age of approximately \combinedage{}.}
\label{fig:ewli_rot_eva}
\end{figure*}

Our analysis indicates that the dominant population of 3D kinematic UMa candidates is well described by a single age of \combinedage{}.
The agreement across all three age–dating approaches is shown in Figure~\ref{fig:ewli_rot_eva}.
The convergence of our independent methods provides strong evidence for a young coeval population within our UMa candidates. The historical spread from 200\,Myr to 1\,Gyr (Table~\ref{tab:uma_ages}) likely results from varying degrees of field star contamination and methodological limitations in earlier studies. 

Our refined age provides essential context for interpreting planetary system evolution, placing UMa in an important evolutionary window where many stars and their planetary systems are transitioning out of their most active phases. We acknowledge that the broader kinematic selection includes a substantial level of contamination. As a result, secondary lines of evidence for youth are critical for confirming true UMa membership among planet hosts. For example, the rotational signature of HD\,63433 lies precisely along the gyrochronological sequence defined by the dominant coeval population \citep{2024AJ....167...54C}, lending strong support to its youth and membership. Other UMa planet-host candidates will similarly require corroborating indicators to establish that they belong to the genuinely young dominant component of the group.

In the case of HD\,63433, a well-constrained age enables more realistic modeling of atmospheric escape rates for its sub-Neptune and Earth-sized planets, which may be evolving from primordial hydrogen envelopes toward secondary atmospheres. Additionally, UMa members occupy the era when debris disks have largely dissipated their gas and are beginning to resemble stable Kuiper belt analogs, providing a useful framework for studying the settling of planetary system architectures.

\subsection{Future Work}
\label{sec:future}
This work leveraged archival data for stellar RVs and lithium equivalent widths. \textit{Gaia} DR3 provides RVs for many of our stars, but there are a number of proper motion candidates that lack RV data. 
In a future paper, we will fill in gaps of measured RVs for some southern hemisphere proper motion candidates with allocated SALT observation time. Additionally, we plan to use SALT spectra to measure lithium equivalent widths, more than doubling the currently available number of lithium data points.

A key component of this ongoing effort is quantifying the contamination rate among kinematic UMa candidates and constructing a catalog of vetted UMa members. 
By combining lithium abundance diagnostics with our ensemble age framework, we can identify stars that satisfy kinematic criteria but exhibit lithium depletion inconsistent with the group's age. 
This will yield a high-fidelity membership list suitable for demographic studies and planet searches.

In this paper, measured rotation periods are based on individual TESS sectors, limiting the reliable rotation periods to ${\sim}$15 days. In the future, with improved methods for stitching together TESS sectors as well as the two longer baseline TESS sectors, we can lessen the bias toward faster rotating stars.
In addition, we plan to search for exoplanets around UMa candidate member stars. Performing a thorough search of TESS light curves for transiting planets in conjunction with a targeted high-precision RV search, we hope to increase the number of young exoplanets with reliable age estimates.

\section{Conclusions}
\label{sec:conclusion}
In this work, we present a comprehensive characterization of the Ursa Major Moving Group, providing context for the planet host HD\,63433 and a foundation for future planetary discoveries within this system.
\begin{itemize}
    \item We constructed a 3D kinematic catalog of UMa candidate members using \textit{Gaia} DR3 astrometry and archival RVs, establishing the largest homogeneously selected sample to date.
    \item We developed an ensemble age-dating framework combining lithium equivalent widths, gyrochronology, and photometric variability, determining a dominant group age of \combinedage{}. Our age estimate is in excellent agreement with the precise interferometric estimate by \cite{Jones15}.
    \item We estimate a ${\sim}40\%$ contamination rate among kinematic candidates using rotational clustering, which is in excellent agreement with the contamination estimate by \cite{dopcke19}.
\end{itemize}

Our results confirm UMa as a benchmark coeval population at \combinedage{}. 
This system is one of the nearest laboratories for studying planetary system evolution at intermediate ages.
However, the substantial contamination rate underscores that single moving group membership criteria are insufficient for reliable characterization. 
We caution against uniformly assigning group ages to all candidate members without multi-dimensional validation. 
As young exoplanet discoveries increasingly rely on host star ages derived from moving group membership, robust vetting approaches like those presented here will be essential for accurate planetary characterization.

\section{Acknowledgments} 
\label{sec:acknowledgments}
Support for this research was provided by the Office of the Vice Chancellor for Research and Graduate Education at the University of Wisconsin--Madison with funding from the Wisconsin Alumni Research Foundation.
AD and RSN express gratitude to the \textit{Peter Livingston Scholars Program}, whose support of undergraduate research provided helpful contributions to this work.
Resources for this project were provided in part by the Wisconsin Center for Origins Research at the
University of Wisconsin--Madison.
This paper includes data collected by the TESS mission, which are publicly available from the Mikulski Archive for Space Telescopes (MAST). Funding for the TESS mission is provided by NASA's Science Mission Directorate.  
This research has made use of the NASA Exoplanet Archive, which is operated by the California Institute of Technology under contract with NASA under the Exoplanet Exploration Program \citep{Akeson2013}.
This work has made use of data from the European Space Agency (ESA) mission \emph{Gaia}\footnote{\url{https://www.cosmos.esa.int/gaia}}, processed by the \emph{Gaia} Data Processing and Analysis Consortium (DPAC)\footnote{\url{https://www.cosmos.esa.int/web/gaia/dpac/consortium}}. 
This research has made use of the VizieR catalogue access tool, CDS, Strasbourg, France. The original description of the VizieR service was published in A\&AS 143, 23. This research has also made use of the SIMBAD database, operated at CDS, Strasbourg, France.
We acknowledge the use of public TOI Release data from pipelines at the TESS Science Office and at the TESS Science Processing Operations Center. This research was achieved using the POLLUX database (pollux.oreme.org) operated at LUPM (Université de Montpellier - CNRS, France) with the support of the PNPS and INSU.
\facilities{\textit{Gaia} DR3 \citep{GaiaDR3}, GALAH DR4 \citep{galahdr4}, Mikulski Archive for Space Telescopes \citep{MAST}, Transiting Exoplanet Survey Satellite \citep{Ricker2015}}

\software{
\texttt{astroquery} \citep{astroquery}, 
astropy \citep{Astropy:2013, Astropy:2018},
\texttt{EAGLES} \citep{eagles},
\texttt{EVA} \citep{Barber23},
\texttt{gyrointerp} \citep{2023ApJ...947L...3B},
\texttt{Lightkurve} \citep{lightkurve}
\texttt{matplotlib} \citep{matplotlib}, 
\texttt{numpy} \citep{Harris:2020}, 
\texttt{PAdova and TRieste Stellar Evolution Code} \citep{parsec2012}, 
\texttt{pandas} \citep{pandas:2023}, 
\texttt{seaborn} \citep{Waskom2021}, 
\texttt{scipy} \citep{Virtanen:2020}, 
\texttt{scikit-learn} \citep{Pedregosa2011},
\texttt{wdwarfdate} \citep{Kiman2022}
}

\bibliography{bibliography}{}
\bibliographystyle{aasjournalv7}

\end{document}